\documentclass[11pt]{article}
\pdfoutput=1
 \usepackage{mciteplus}
 \usepackage{tikz}
 \usepackage{color}
 \definecolor{darkblue}{rgb}{0.1,0.1,.7}
 \usepackage[colorlinks, linkcolor=darkblue, citecolor=darkblue, urlcolor=darkblue, linktocpage,hyperfootnotes=false]{hyperref} 
\usepackage{epsfig}
\usepackage{graphicx}
\usepackage{cite}
\usepackage{amsfonts}
\usepackage{amssymb}
\usepackage{bm}
\usepackage{latexsym}
\usepackage{amsmath}
\def\bq{\begin{quote}}
\def\eq{\end{quote}}
 at 10truept

\newcommand{\calo}{{\cal O}}

\newcommand{\calu}{{\cal U}}

\newcommand{\beq}{\begin{equation}}
\newcommand{\eeq}{\end{equation}}
\newcommand{\beqa}{\begin{eqnarray}}
\newcommand{\eeqa}{\end{eqnarray}}
\newcommand{\bea}{\begin{eqnarray}}
\newcommand{\eea}{\end{eqnarray}}

\newcommand{\hf}{\frac{1}{2}}

\def\roughly#1{\raise.3ex\hbox{$#1$\kern-.75em\lower1ex\hbox{$\sim$}}}

\begin{document}

\thispagestyle{empty}
\begin{titlepage}
 \setcounter{page}{0}
  \bigskip

  \bigskip\bigskip

  \bigskip

\begin{center}
{\Large \bf {On the questions of asymptotic recoverability of information and subsystems in quantum gravity}}
    \bigskip
\bigskip
\end{center}

  \begin{center}

 \rm {Steven B. Giddings\footnote{\texttt{giddings@ucsb.edu}} }
  \bigskip \rm
\bigskip

{Department of Physics, University of California, Santa Barbara, CA 93106, USA}  \\
\rm

  \bigskip \rm
\bigskip
 
\rm

\bigskip
\bigskip

  \end{center}

\vspace{3cm}
  \begin{abstract}
  A longstanding question in quantum gravity regards the localization of quantum information; one way to formulate this question is to ask how subsystems can be defined in quantum-gravitational systems.  The gauge symmetry and necessity of solving the constraints appear to imply that the answers to this question here are different than in finite quantum systems, or in local quantum field theory.  Specifically, the constraints can be solved by providing a ``gravitational dressing" for the underlying field-theory operators, but this modifies their locality properties.  It has been argued that holography itself may be explained through this role of the gauge symmetry and constraints, at the nonperturbative level, but there are also subtleties in constructing a holographic map in this approach.  There are also claims that holography is implied even by perturbative solution of the constraints.   This short note provides further examination of these questions, and in particular investigates to what extent perturbative or nonperturbative solution of the constraints implies that information na\"\i vely thought to be localized can be recovered by asymptotic measurements, and the relevance of this in defining subsystems.  In the leading perturbative case, the relevant effects are seen to be exponentially suppressed.  These questions are, for example, important in sharply characterizing the unitarity problem for black holes.

 \medskip
  \noindent
  \end{abstract}
\bigskip \bigskip \bigskip 

  \end{titlepage}

%

It has long been believed that localization of information in quantum gravity may behave differently than in local quantum field theory (LQFT).  Certainly this has been one of the themes of the proposal of holography \cite{tHooholo,Sussholo,Malda}, and is also motivated by consideration of the diffeomorphism gauge symmetry of classical and perturbative quantum general relativity (GR).  In trying to understand this localization, there has been increasing focus on the role of the constraints of perturbative GR and its possible nonperturbative generalization, and on  properties of the gauge-invariant states and operators annihilated by the constraints.  A particularly important question to answer, underlying a description of many information-theoretic aspects of quantum gravity, is in what sense one can precisely define {\it quantum subsystems} in the theory.  For finite or locally finite quantum systems, subsystems are defined in terms of factorization of Hilbert spaces, and for LQFT systems they are defined in terms of commuting subalgebras of observables.  But in perturbative GR, the situation is more subtle; the constraints can be solved by dressing underlying states or operators of LQFT\cite{SGalg,DoGi1,DoGi2}\footnote{Earlier related work includes \cite{Heem} and  \cite{KaLigrav}.   The former exhibited nontrivial commutators arising from the constraints but did not describe the dressed operators; the latter was focused on finding bulk operators that {\it commute}, and did not exhibit  the bulk dressing described below.}  (for some further development see \cite{DoGi3,DoGi4,GiKi,SGsplit}), and this dressing obstructs na\"\i ve extensions of such factorization or subalgebras.

Indeed, perhaps the leading candidate for an explanation of holography is that it arises precisely from the gauge symmetry and constraints of gravity; arguments in this direction have been given in \cite{MaroUH,Marothought,MaroholoNS,JacoBU}.  There are certain subtleties\cite{JaNg,HoUn} in these arguments, so this remains an actively investigated question.

In trying to make the question of localization of information more concrete, a specific goal is to understand how 
to define subsystems in gravity, since in other quantum systems such a definition is at the basis of explaining how information can be localized.  This question has in particular been preliminarily discussed in gravity, taking into account the preceding considerations, in \cite{SGalg,DoGi4,SGsplit}.  It is also an important question in addressing concrete problems for quantum gravity such as that of explaining the reconciliation of black hole evolution with quantum mechanics; for example the assumption of the (approximate) existence of subsystems is one of the underlying assumptions in a ``black hole theorem\cite{BHthm}" that sharply constrains the possible avenues for unitary evolution.

This short paper will examine some aspects and subtleties regarding the question of defining subsystems, in view of the constraints and properties of the dressing.  One way to begin to address these is to ask a question: to what extent is information ``recoverable" asymptotically in gravity, in situations where it would not be recoverable in LQFT, by making observations or performing experiments in an asymptotic region far from a region where a na\"\i ve field theory analysis would tell one it is localized?

The formal arguments for holography\cite{MaroUH} do suggest that information is asymptotically recoverable, although these do ultimately seem to be based on assuming a non-perturbative solution of the constraints, among other subtleties\cite{HoUn}.  A related argument was given in \cite{DoGi3} (see below), that given a full solution of the constraints, one can use the fact that the translation operator is a boundary operator to translate a state to an asymptotic region, where it can be measured.  
There are also recent claims\cite{CGPR} that based on perturbative construction of dressings that solve the constraints, even at the perturbative level information is recoverable asymptotically.  This paper will specifically explore some aspects of this perturbative vs. nonperturbative recoverability of information.

As a test case, consider perturbative quantization of two scalar fields, $\phi_a$, $a=1,2$, coupled to gravity, preserving the global symmetry distinguishing them.\footnote{There are arguments that such global symmetries are spoiled nonperturbatively in gravity, but this is not obviously true in the 
perturbative setting, and so let's remain agnostic on this point.}  We will begin with the case of a spacetime $M$ with general asymptotics ({\it e.g.} Minkowski or AdS), with simplifying restrictions in later examples.

To test recoverability of information, consider the two states
\beq
\label{Jstate}
|J\rangle_a= e^{-i\int J(x)\phi_a(x)} |0\rangle = {\cal U}_{Ja}|0\rangle\ ,
\eeq
where $J(x)$ is a source function with compact support in some neighborhood $U$ of $M$, or consider the dressed version of these states in the full gravitational theory.

First, in LQFT, we know that information about which state we choose  is not recoverable at spacelike separation to $U$, through any  measurements.  To review the argument, if we consider measuring a correlator of some collection $\calo_A$ of operators spacelike to $U$, then
\beq
{}_a\langle J| \prod_A {\cal O}_A |J\rangle_a = \langle 0|\prod_A {\cal O}_A |0\rangle\ ,
\eeq
due to the operators $\calo_A$ commuting with $\phi(x)$ at spacelike separation.  Thus such observations cannot distinguish the states $|J\rangle_a$, and cannot even distinguish them from vacuum.

The situation changes in gravity, once we solve the constraints to determine a gravitational dressing for the state; this is because in general this dressing must extend to infinity\cite{DoGi2}, affecting measurements there.  In particular, if we consider a collection of Poincar\'e charges $Q_\alpha$, 
we can use asymptotic measurements to determine ${}_a\langle J|\prod_\alpha Q_\alpha |J\rangle_a$.   Such measurements {\it do} distinguish $|J\rangle_a$ from vacuum, but do not distinguish the $|J\rangle_a$, with $a=1,2$, from each other.

To see how this works, minimally couple the fields $\phi_a$ to gravity with the Einstein action.  With a choice of time slicing, the constraints take the form
\beq
C_\mu(x)=  \frac{1}{8\pi G} G_{0\mu}(x)-T_{0\mu}(x) =0\ ,
\eeq 
where in the AdS case $T$ includes the
 cosmological term.  The quantum version of these generalize the Wheeler-DeWitt equation, which corresponds to the $\mu=0$ component.
These may be solved by working with quantum deformations of a fixed background metric $g_{\mu\nu}$
\beq
\tilde g_{\mu\nu}= g_{\mu\nu} + \kappa h_{\mu\nu}, 
\eeq
with $\kappa^2=32\pi G$.  
At the quantum level, diffeomorphism-invariant operators are those commuting with the constraints, and states 
solving the constraints are annihilated by the $C_\mu$.\footnote{As in Gupta-Bleuler quantization\cite{Gupt}, one actually requires that only half of the $C_\mu$ annihilate the state, as noted in \cite{DoGi4}.}  For example, the states \eqref{Jstate} may be promoted to such solutions,
\beq
|J\rangle_a = {\cal U}_{Ja}|0\rangle \rightarrow |\widehat J\rangle_a = \hat \calu_{Ja}[\phi,h]|0\rangle\ ,
\eeq
where the operators  $\hat\calu_{Ja}[\phi,h]$ now depend also on the metric perturbation, and are to be determined.  

Finding complete solutions is a challenging problem, but perturbative solutions to linear order in $\kappa$ have been studied in \cite{SGalg,DoGi1,DoGi2,DoGi3,DoGi4,QGQF,SGsplit}.\footnote{Ref.~\cite{CGPR} likewise described perturbative solution of the constraints, but failed to note the direct relation of their perturbative analysis to this construction; one can see that the dressing constructions are in many respects simpler than their analysis. Ref.~\cite{GKPRR} also discussed implications of dressing for the ``islands" discussion \cite{AHMST,PSSY}.}  
These take the form
\beq\label{dressj}
 |\widehat J\rangle_a \simeq e^{i\int d^3x V^\mu(x)T_{0\mu}(x)} |J\rangle_a\ ,
 \eeq
 where $V^\mu(x)$ are functionals of $h$ that are linear at first order in $\kappa$, and are also a function of position $x$.  Under a diffeomorphism transforming $\delta h_{\mu\nu} = -\partial_\mu\xi_\nu - \partial_\nu\xi_\mu +\calo(\kappa)$, these vary according to the key relation
 \beq\label{keyreln}
 \delta V^\mu(x) = \kappa\xi^\mu(x)\ , 
 \eeq
 but the $V^\mu$ are not uniquely fixed, and indeed differences correspond to adding a source free radiation field $h$ to a given dressing.\footnote{This corresponds to the dependence on $h^{TT}$ described in \cite{CGPR} .}  Working about Minkowski space, a broad class of these takes the form\cite{SGsplit} (specified, say, at $t=0$)
 \beq
 V_\mu(0,\vec x)=\int  d^3 x' \check h^{ij}_{\vec x}(\vec x')  \gamma_{\mu,ij}(0,\vec x')
 \eeq
 with 
 \beq
 \gamma_{\mu,ij}= \frac{\kappa}{2} (\partial_i h_{\mu j} + \partial_j h_{\mu i} - \partial_\mu h_{ij})
  \eeq
 the linearized Christoffel symbol, and with a classical tensor field $\check h^{ij}_{\vec x}$ satisfying
 \beq
 \partial_i\partial_j \check h^{ij}_{\vec x}(\vec x') = -\delta^3(\vec x' -\vec x)\ .
 \eeq
 A special case is the line dressing of \cite{DoGi1,QGQF},
 \beq
 V_\mu^L(x,y)= -\frac{\kappa}{2} \int_y^x dx^{\prime\nu} \left\{ h_{\mu\nu}(x') - 
 \int_y^{x'} dx^{\prime\prime\lambda} \left[ \partial_\mu h_{\nu\lambda}(x'') - \partial_\nu h_{\mu\lambda}(x'')\right]\right\}\ ,
 \eeq
with $y=\infty$, giving a gravitational line running to infinity.

Properties of the dressing can be  illustrated by calculating the expectation value of the leading order metric perturbation, using \eqref{dressj}
\beq\label{hexp}
{}_a\langle \widehat J|h_{\mu\nu} (z) |\widehat J\rangle_a \simeq  i\int d^3x [h_{\mu\nu}(z),V^\lambda(x)]\  {}_a\langle J|T_{0\lambda}(x) |J\rangle_a\ .
\eeq
 This is clearly independent of $a$.  The different choices of dressings lead to different values for the commutator and hence for the expectation value.  However, much of this freedom arises precisely from adding the different free radiation fields on top of a dressing necessary to solve the constraints, {\it e.g.} by shifting $\check h_{\vec x}^{ij}$ by a source-free solution.  Any such dressing and expectation value must however give the correct values for the asymptotically-measured Poincar\'e charges, $P_\mu$, $M_{\mu\nu}$.  
 
 To see that this is the {\it only} necessary correlation with the choice of matter state,\footnote{Perturbative gravitons can also be included.} one can use the ``standard dressing" construction as in \cite{DoGi4,SGsplit}.  Suppose we choose a point $y\in U$ and {\it some} dressing $V^\mu_S(y)$ satisfying the key relation \eqref{keyreln}.  Then, for general $x\in U$, define the dressing
 \beq
 V^\mu(x) = V_L^\mu(x,y) + V^\mu_S(y) + \hf (x-y)_\nu \left[\partial^\nu V^\mu_S(y) - \partial^\mu V^\nu_S(y)\right]\ .
 \eeq
 Given the commutator
 \beq
  i[h_{\mu\nu}(z),V_S^\lambda(y)] = -h_{\mu\nu}^{S\lambda}(z,y)\ ,
  \eeq
 for $z$ outside $U$, the commutator in  \eqref{hexp} becomes
 \beq
 i[h_{\mu\nu}(z),V^\lambda(x)] = -h_{\mu\nu}^{S\lambda}(z,y) - \hf (x-y)_\sigma \left[ \partial_y^\sigma h^{S\lambda}_{\mu\nu}(z,y) -  \partial_y^\lambda h^{S\sigma}_{S\mu\nu}(z,y)\right]\ , 
 \eeq
 since the metric perturbation outside $U$ commutes with the line dressing connecting $x$ and $y$.  Then, the expectation value \eqref{hexp} becomes
 \beq\label{hexpr}
 {}_a\langle \widehat J|h_{\mu\nu} (z) |\widehat J\rangle_a \simeq  h_{\mu\nu}^{S\lambda}(z,y)\ {}_a\langle J|P_\lambda |J\rangle_a + \hf \partial_y^\sigma h^{S\lambda}_{\mu\nu}(z,y)\ 
 {}_a\langle J|M_{\sigma \lambda} |J\rangle_a\ ,
 \eeq
 in terms of the total Poincar\'e charges $P_\mu$ and $M_{\mu\nu}$, with the latter defined with respect to an origin at $y$.  This metric expectation value thus depends on the choice of point $y$ and standard dressing $V^\mu_S$, and on the total Poincar\'e charges of the matter configuration.
 
 If one instead considers an n-point function ${}_a\langle \widehat J|h_{\mu_1\nu_1} (z_1)\cdots h_{\mu_n\nu_n} (z_n) |\widehat J\rangle_a$, then commuting the exponential in \eqref{dressj} through the $h_{\mu\nu}(z_A)$ yields a leading order contribution\cite{SGsplit} from a product of terms like \eqref{hexpr},
 \beq\label{npt}
 {}_a\langle \widehat J|h_{\mu_1\nu_1} (z_1)\cdots h_{\mu_n\nu_n} (z_n) |\widehat J\rangle_a 
 = {}_a\langle J| \prod_{A=1}^n \left[ h_{\mu_A\nu_A}^{S\lambda_A}(z_A,y)P_{\lambda_A} +
  \hf \partial_y^{\sigma_A} h^{S\lambda_A}_{\mu_A\nu_A}(z_A,y) M_{\sigma_A\lambda_A}  \right] |J\rangle_a\ ,
 \eeq
 now depending on the state $|J\rangle_a$ only through moments of its Poincar\'e charges.  
 A special case of these expressions, where one computes soft charges from integrating the asymptotic $h_{\mu_A\nu_A}(z_A)$, was in \cite{SGsplit} argued to show that the only required correlation of the soft charges with the state is likewise through the moments of the Poincar\'e charges.
  
While this shows that at the perturbative level measurements of the metric asymptotics are only sensitive to Poincar\'e charges, arguments have been given that more general asymptotic measurements can be used to determine the state given the full, rather than leading order, dressing.  One is the argument of \cite{MaroUH} (for further discussion see \cite{JacoBU} and \cite{HoUn}), and an even simpler one was given in \cite{DoGi3}.  These rely on the fact that in gravity the total momentum can be written as
\beq
P_\mu = P_\mu^{ADM}[h(\infty)] + \int d^3x\  C_\mu(x)\ ,
\eeq
with $P_\mu^{ADM}$ the ADM expressions given in terms of surface integrals at infinity,
and so for a solution of the constraints is given by just these ADM terms.  So, following \cite{DoGi3} we can consider the expectation value of the asymptotic operators
\beq\label{NPtrans}
{}_a\langle\widehat J|\phi_b(y)\, e^{i P_i^{ADM} c^i} |\widehat J\rangle_a \ 
\eeq
where $y$ is now in the asymptotic region, and $c^i$ describes a large translation that moves the support of $J$ into the asymptotic region overlapping $y$. 
For a solution of the constraints the exponential in \eqref{NPtrans} is equivalent to one with the  full momentum $P_i$, and this
has the effect of translating the state $|\widehat J\rangle_a$ to this asymptotic region, where the operator $\phi_b(y)$ can then ``register" the state with a result $\propto\delta_{ab}$ and distinguish the two possible states \eqref{Jstate}.  This, thus, describes asymptotic recoverability of information, for a full nonperturbative solution of the constraints.

While these arguments have been illustrated in a Minkowski background, one expects them to straightforwardly generalize to the context of anti de Sitter space, with the same structure, using constructions of AdS dressings solving the constraints like those given in \cite{GiKi}.  Thus, for example, one can construct a formal argument that by measuring combinations of matter and gravitational operators analogous to \eqref{NPtrans} at the AdS boundary, given a nonperturbative solution of the constraints, one can asymptotically recover information about the state.  

Refs.~\cite{LPRS,CGPR} have recently argued that this asymptotic accessibility of information also extends to the perturbative context.  
Specifically, 
suppose we have perturbatively solved the constraints, to determine a dressing.  Working again with the example of a Minkowski background, and generalizing eq.~\eqref{npt}, or the preceding argument, one might expect that asymptotic measurements allow determination of the 
correlators
\beq\label{Qcorr}
{}_a\langle J| \prod_A {\cal O}_A\prod_\alpha Q_\alpha |J\rangle_a\ 
\eeq
where again $Q_\alpha$ are Poincar\'e charges, and $\calo_A$ are asymptotic operators, {\it e.g.} corresponding to matter operators.
Let's suppose that we can indeed measure such correlators via asymptotic measurements.  Does that allow us to determine the difference between the $|J\rangle_a$?  We emphasize that, once we have taken into account the use of the dressing in providing the means to make the asymptotic measurements, \eqref{Qcorr} is then regarded as an expression on a fixed background metric, {\it e.g.} AdS or Minkowski.
Once again using the definition \eqref{Jstate} of the state, and the commutativity of the asymptotic operators $\calo_A$ with $\calu_{Ji}$, the correlator \eqref{Qcorr} becomes
\beq\label{Qcorr2}
{}_a\langle J| \prod_A {\cal O}_A\prod_\alpha Q_\alpha |J\rangle_a  = \langle0| \prod_A \calo_A \prod_\alpha \left(\calu_{Ja}^\dagger Q_\alpha \calu_{Ja}\right)|0\rangle\ .
\eeq

To understand the dependence of the latter expression on the state $|J\rangle_a$, consider the special case where the charge is the hamiltonian, as was considered in \cite{CGPR}.  In this case, for free scalar fields, one finds
\beq\label{Hconj}
\calu_{Ja}^\dagger H \calu_{Ja} = H+ \int d^3x \left(\dot \phi_J \dot \phi_a+ \vec\partial \phi_J \cdot  \vec\partial \phi_a\right) + E_J\ ,
\eeq
where $\phi_J$ is the classical solution with source $J$, 
\beq
\Box\phi_J = J(x)\ ,
\eeq
which has support only in the future lightcone of $U$, and $E_J$ is the energy of this solution.  

We now find that there can be dependence of the correlators \eqref{Qcorr}, \eqref{Qcorr2} on the choice of $a$. Once again, to take a simple example, consider the correlator
\beq
{}_a\langle J| \phi_b(y) H |J\rangle_a
\eeq
where $y$ is in a region asymptotically far from $U$.  The preceding steps then yield
\beq\label{simpcorr}
{}_a\langle J| \phi_b(y) H |J\rangle_a = \langle0| \phi_b(y) \int d^3x \left(\dot \phi_J \dot \phi_a+ \vec\partial \phi_J \cdot  \vec\partial \phi_a\right) |0\rangle\ ,
\eeq
which is proportional to $\delta_{ab}$, seemingly registering the difference between the two states \eqref{Jstate}.

However, the result \eqref{simpcorr} is exponentially small\cite{BHthm}.  Specifically, let $m$ be the mass of the fields $\phi_a$, and let $y$ be spatially separated from  $U$ with distance $L$, in the Minkowski example.  Since $\phi_J$ only has support in in the future lightcone of $U$, we have
\beq
{}_a\langle J| \phi_b(y) H |J\rangle_a \sim \delta_{ab}\, e^{-mL}\ ,
\eeq
The recoverable distinction between the states is thus exponentially tiny in the asymptotic separation\cite{BHthm}, and vanishes at infinity.  From the expressions \eqref{Qcorr2} and \eqref{Hconj}, we clearly expect similar exponential suppression for more general correlators.

Ref.~\cite{CGPR} has alternately suggested using arbitrary powers of $H$ in \eqref{Qcorr2} to construct energy projectors.  This has a similar result. Suppose we likewise assume that it is possible to make boundary measurements of such projected correlators.  An example would be
\beq
{}_a\langle J|\phi_b(y) |0\rangle\langle 0|J\rangle_a\ .
\eeq
But, such a correlator, while proportional to $\delta_{ab}$, is again exponentially small in $mL$.

It is an interesting property of gravity that there is such an in-principle asymptotic distinction between the states $|J\rangle_a$ also at the perturbative level discussed here.  
One way of describing what has happened is that access to measurements of local operators {\it and} Poincar\'e charges has given sensitivity to the small field correlations between the field in a neighborhood, and that in its complement.  
What is not clear is that such small effects lead  to a meaningful
perturbative recoverability of the information about the identity of the state by asymptotic measurements.

For one thing, if one performs asymptotic measurements, {\it e.g.} by interacting or scattering of some system with the asymptotic gravitational field, one needs exponential sensitivity to register an effect with a probability that is not exponentially suppressed.  This does not appear to indicate that the information is well recoverable asymptotically. Likewise, for example, transfer of entanglement associated with such information from the original system to a measuring system would be exponentially suppressed.

Moreover, the argument that the correlators 
\eqref{Qcorr} could be determined via asymptotic measurements relied on solving the constraints to determine the gravitationally-dressed state.  If one only has a perturbative construction of the dressing solving the constraints, at some order, it is not clear that it will lead to precise enough determination of the correlators to distinguish $\exp\{-mL\}$ from zero.  The question of perturbative solution of the constraints to determine the higher-order dressings, generalizing\cite{DoGi1}, and its accuracy and role in such arguments, is left to work in progress.

The preceding arguments are amplified by another argument\cite{BHthm} that, in the general situation, one  needs a nonperturbative solution of the constraints if one hopes 
 to recover information in asymptotic regions:  the states \eqref{Jstate} could differ by operators analogous to $\calu_{Ja}$ that act only within a black hole.  This also suggests the need for a nonperturbative dressing, although one could try to formally run the preceding argument by constructing perturbative dressings, like those studied in \cite{GiWe}, on the background geometry of the black hole.
 
 In short, if there are indeed exponentially tiny effects by which information is delocalized, that leaves the question regarding to what extent one can still define an approximate notion of subsystem, in which it is for example dominantly localized.  The preceding arguments suggest that standard measurements or scattering experiments don't have access to such information.
 
 We can specifically return  to the question of black holes, and to what extent a black hole is a subsystem, in which for example information can be localized.  A black hole {\it is} such a system in the LQFT approximation, and that localization lies at the center of the problem of unitarity.  It is certainly true that the small effects we have discussed make this picture more subtle, and in particular for example raise an additional subtlety in defining entropies of black holes or other candidate subsystems in gravity, but that is not clearly yet the resolution of the unitarity problem advocated in recent work (see, {\it e.g.}, \cite{CGPR}).  
 
 Of course, the observation that 
 effective couplings 
 of the black hole state to its environment\cite{NVU,BHQU}  that are exponentially small in the black hole entropy 
 can in principle restore unitarity is also suggestive, in the context of discussing an exponentially small delocalization of quantum information.  But, it remains to be seen whether these effects are connected, and it is not clear that the latter provide the modifications to evolution necessary for unitarity. It would be very useful to more clearly understand the (approximate) localization of information in quantum gravity, and also the full evolution of that information, for which a nonperturbative completion may well be needed.  Again the formal arguments for holography of \cite{MaroUH, DoGi2} appear to rely on a nonperturbative solution of the constraints, which is tantamount to starting with  a nonperturbative description of evolution\cite{HoUn}, suggesting that this doesn't lead to a direct solution to the problem.

\vskip.3in
\noindent{\bf Acknowledgements} 
 
This material is based upon work supported in part by the U.S. Department of Energy, Office of Science, under Award Number {DE-SC}0011702, and by Heising-Simons Foundation grant \#2021-2819.  I thank T. Jacobson for asking questions prompting me to elaborate these observations.

\mciteSetMidEndSepPunct{}{\ifmciteBstWouldAddEndPunct.\else\fi}{\relax}
\bibliographystyle{utphys}
\bibliography{subsyst}{}

\providecommand{\href}[2]{#2}\begingroup\raggedright\begin{thebibliography}{10}

\bibitem{tHooholo}
G.~'t~Hooft, ``{Dimensional reduction in quantum gravity},'' {\em Conf. Proc.
  C} {\bfseries 930308} (1993) 284--296,
  \href{http://arxiv.org/abs/gr-qc/9310026}{{\ttfamily arXiv:gr-qc/9310026}}.

\bibitem{Sussholo}
L.~Susskind, ``{The World as a hologram},''
  \href{http://dx.doi.org/10.1063/1.531249}{{\em J. Math. Phys.} {\bfseries 36}
  (1995) 6377--6396}, \href{http://arxiv.org/abs/hep-th/9409089}{{\ttfamily
  arXiv:hep-th/9409089}}.

\bibitem{Malda}
J.~M. Maldacena, ``{The Large N limit of superconformal field theories and
  supergravity},'' \href{http://dx.doi.org/10.1023/A:1026654312961}{{\em Adv.
  Theor. Math. Phys.} {\bfseries 2} (1998) 231--252},
  \href{http://arxiv.org/abs/hep-th/9711200}{{\ttfamily arXiv:hep-th/9711200}}.

\bibitem{SGalg}
S.~B. Giddings, ``{Hilbert space structure in quantum gravity: an algebraic
  perspective},'' \href{http://dx.doi.org/10.1007/JHEP12(2015)099}{{\em JHEP}
  {\bfseries 12} (2015) 099}, \href{http://arxiv.org/abs/1503.08207}{{\ttfamily
  arXiv:1503.08207 [hep-th]}}.

\bibitem{DoGi1}
W.~Donnelly and S.~B. Giddings, ``{Diffeomorphism-invariant observables and
  their nonlocal algebra},''
  \href{http://dx.doi.org/10.1103/PhysRevD.93.024030}{{\em Phys. Rev. D}
  {\bfseries 93} no.~2, (2016) 024030},
  \href{http://arxiv.org/abs/1507.07921}{{\ttfamily arXiv:1507.07921
  [hep-th]}}. [Erratum: Phys.Rev.D 94, 029903 (2016)].

\bibitem{DoGi2}
W.~Donnelly and S.~B. Giddings, ``{Observables, gravitational dressing, and
  obstructions to locality and subsystems},''
  \href{http://dx.doi.org/10.1103/PhysRevD.94.104038}{{\em Phys. Rev. D}
  {\bfseries 94} no.~10, (2016) 104038},
  \href{http://arxiv.org/abs/1607.01025}{{\ttfamily arXiv:1607.01025
  [hep-th]}}.

\bibitem{Heem}
I.~Heemskerk, ``{Construction of Bulk Fields with Gauge Redundancy},''
  \href{http://dx.doi.org/10.1007/JHEP09(2012)106}{{\em JHEP} {\bfseries 09}
  (2012) 106}, \href{http://arxiv.org/abs/1201.3666}{{\ttfamily arXiv:1201.3666
  [hep-th]}}.

\bibitem{KaLigrav}
D.~Kabat and G.~Lifschytz, ``{Decoding the hologram: Scalar fields interacting
  with gravity},'' \href{http://dx.doi.org/10.1103/PhysRevD.89.066010}{{\em
  Phys. Rev. D} {\bfseries 89} no.~6, (2014) 066010},
  \href{http://arxiv.org/abs/1311.3020}{{\ttfamily arXiv:1311.3020 [hep-th]}}.

\bibitem{DoGi3}
W.~Donnelly and S.~B. Giddings, ``{How is quantum information localized in
  gravity?},'' \href{http://dx.doi.org/10.1103/PhysRevD.96.086013}{{\em Phys.
  Rev. D} {\bfseries 96} no.~8, (2017) 086013},
  \href{http://arxiv.org/abs/1706.03104}{{\ttfamily arXiv:1706.03104
  [hep-th]}}.

\bibitem{DoGi4}
W.~Donnelly and S.~B. Giddings, ``{Gravitational splitting at first order:
  Quantum information localization in gravity},''
  \href{http://dx.doi.org/10.1103/PhysRevD.98.086006}{{\em Phys. Rev. D}
  {\bfseries 98} no.~8, (2018) 086006},
  \href{http://arxiv.org/abs/1805.11095}{{\ttfamily arXiv:1805.11095
  [hep-th]}}.

\bibitem{GiKi}
S.~B. Giddings and A.~Kinsella, ``{Gauge-invariant observables, gravitational
  dressings, and holography in AdS},''
  \href{http://dx.doi.org/10.1007/JHEP11(2018)074}{{\em JHEP} {\bfseries 11}
  (2018) 074}, \href{http://arxiv.org/abs/1802.01602}{{\ttfamily
  arXiv:1802.01602 [hep-th]}}.

\bibitem{SGsplit}
S.~B. Giddings, ``{Gravitational dressing, soft charges, and perturbative
  gravitational splitting},''
  \href{http://dx.doi.org/10.1103/PhysRevD.100.126001}{{\em Phys. Rev. D}
  {\bfseries 100} no.~12, (2019) 126001},
  \href{http://arxiv.org/abs/1903.06160}{{\ttfamily arXiv:1903.06160
  [hep-th]}}.

\bibitem{MaroUH}
D.~Marolf, ``{Unitarity and Holography in Gravitational Physics},''
  \href{http://dx.doi.org/10.1103/PhysRevD.79.044010}{{\em Phys. Rev. D}
  {\bfseries 79} (2009) 044010},
  \href{http://arxiv.org/abs/0808.2842}{{\ttfamily arXiv:0808.2842 [gr-qc]}}.

\bibitem{Marothought}
D.~Marolf, ``{Holographic Thought Experiments},''
  \href{http://dx.doi.org/10.1103/PhysRevD.79.024029}{{\em Phys. Rev. D}
  {\bfseries 79} (2009) 024029},
  \href{http://arxiv.org/abs/0808.2845}{{\ttfamily arXiv:0808.2845 [gr-qc]}}.

\bibitem{MaroholoNS}
D.~Marolf, ``{Holography without strings?},''
  \href{http://dx.doi.org/10.1088/0264-9381/31/1/015008}{{\em Class. Quant.
  Grav.} {\bfseries 31} (2014) 015008},
  \href{http://arxiv.org/abs/1308.1977}{{\ttfamily arXiv:1308.1977 [hep-th]}}.

\bibitem{JacoBU}
T.~Jacobson, ``{Boundary unitarity and the black hole information paradox},''
  \href{http://dx.doi.org/10.1142/S0218271813420029}{{\em Int. J. Mod. Phys. D}
  {\bfseries 22} (2013) 1342002},
  \href{http://arxiv.org/abs/1212.6944}{{\ttfamily arXiv:1212.6944 [hep-th]}}.

\bibitem{JaNg}
T.~Jacobson and P.~Nguyen, ``{Diffeomorphism invariance and the black hole
  information paradox},''
  \href{http://dx.doi.org/10.1103/PhysRevD.100.046002}{{\em Phys. Rev. D}
  {\bfseries 100} no.~4, (2019) 046002},
  \href{http://arxiv.org/abs/1904.04434}{{\ttfamily arXiv:1904.04434 [gr-qc]}}.

\bibitem{HoUn}
S.~B. Giddings, ``{Holography and unitarity},''
  \href{http://dx.doi.org/10.1007/JHEP11(2020)056}{{\em JHEP} {\bfseries 11}
  (2020) 056}, \href{http://arxiv.org/abs/2004.07843}{{\ttfamily
  arXiv:2004.07843 [hep-th]}}.

\bibitem{BHthm}
S.~B. Giddings, ``{A `black hole theorem,' and its implications},''
  \href{http://arxiv.org/abs/2110.10690}{{\ttfamily arXiv:2110.10690
  [hep-th]}}.

\bibitem{CGPR}
C.~Chowdhury, V.~Godet, O.~Papadoulaki, and S.~Raju, ``{Holography from the
  Wheeler-DeWitt equation},'' \href{http://arxiv.org/abs/2107.14802}{{\ttfamily
  arXiv:2107.14802 [hep-th]}}.

\bibitem{Gupt}
S.~N. Gupta, ``Quantization of Einstein's Gravitational Field: Linear
  Approximation,'' {\em Proc. Phys. Soc.} {\bfseries 65} no.~3, (1952) 161.

\bibitem{QGQF}
S.~B. Giddings, ``{Quantum gravity: a quantum-first approach},''
  \href{http://dx.doi.org/10.31526/LHEP.3.2018.01}{{\em LHEP} {\bfseries 1}
  no.~3, (2018) 1--3}, \href{http://arxiv.org/abs/1805.06900}{{\ttfamily
  arXiv:1805.06900 [hep-th]}}.

\bibitem{GKPRR}
H.~Geng, A.~Karch, C.~Perez-Pardavila, S.~Raju, L.~Randall, M.~Riojas, and
  S.~Shashi, ``{Inconsistency of Islands in Theories with Long-Range
  Gravity},'' \href{http://arxiv.org/abs/2107.03390}{{\ttfamily
  arXiv:2107.03390 [hep-th]}}.

\bibitem{AHMST}
A.~Almheiri, T.~Hartman, J.~Maldacena, E.~Shaghoulian, and A.~Tajdini,
  ``{Replica Wormholes and the Entropy of Hawking Radiation},''
  \href{http://dx.doi.org/10.1007/JHEP05(2020)013}{{\em JHEP} {\bfseries 05}
  (2020) 013}, \href{http://arxiv.org/abs/1911.12333}{{\ttfamily
  arXiv:1911.12333 [hep-th]}}.

\bibitem{PSSY}
G.~Penington, S.~H. Shenker, D.~Stanford, and Z.~Yang, ``{Replica wormholes and
  the black hole interior},'' \href{http://arxiv.org/abs/1911.11977}{{\ttfamily
  arXiv:1911.11977 [hep-th]}}.

\bibitem{LPRS}
A.~Laddha, S.~G. Prabhu, S.~Raju, and P.~Shrivastava, ``{The Holographic Nature
  of Null Infinity},''
  \href{http://dx.doi.org/10.21468/SciPostPhys.10.2.041}{{\em SciPost Phys.}
  {\bfseries 10} no.~2, (2021) 041},
  \href{http://arxiv.org/abs/2002.02448}{{\ttfamily arXiv:2002.02448
  [hep-th]}}.

\bibitem{GiWe}
S.~Giddings and S.~Weinberg, ``{Gauge-invariant observables in gravity and
  electromagnetism: black hole backgrounds and null dressings},''
  \href{http://dx.doi.org/10.1103/PhysRevD.102.026010}{{\em Phys. Rev. D}
  {\bfseries 102} no.~2, (2020) 026010},
  \href{http://arxiv.org/abs/1911.09115}{{\ttfamily arXiv:1911.09115
  [hep-th]}}.

\bibitem{NVU}
S.~B. Giddings, ``{Nonviolent unitarization: basic postulates to soft quantum
  structure of black holes},''
  \href{http://dx.doi.org/10.1007/JHEP12(2017)047}{{\em JHEP} {\bfseries 12}
  (2017) 047}, \href{http://arxiv.org/abs/1701.08765}{{\ttfamily
  arXiv:1701.08765 [hep-th]}}.

\bibitem{BHQU}
S.~B. Giddings, ``{Black holes in the quantum universe},''
  \href{http://dx.doi.org/10.1098/rsta.2019.0029}{{\em Phil. Trans. Roy. Soc.
  Lond. A} {\bfseries 377} no.~2161, (2019) 20190029},
  \href{http://arxiv.org/abs/1905.08807}{{\ttfamily arXiv:1905.08807
  [hep-th]}}.

\end{thebibliography}\endgroup

\end{document}